  \documentclass{cambridge6A}
  \usepackage[numbers]{natbib}

  \usepackage{rotating}
  \usepackage{floatpag}
  \rotfloatpagestyle{empty}

  \usepackage{amsthm}
  \usepackage{graphicx}

\def\mathnew{\mathsurround=0pt}
\def\simov#1#2{\lower .5pt\vbox{\baselineskip0pt \lineskip-.5pt
       \ialign{$\mathnew#1\hfil##\hfil$\crcr#2\crcr\sim\crcr}}}

\def\gtrsim{\mathrel{\mathpalette\simov >}}

\def\lesssim{\mathrel{\mathpalette\simov <}}

\def\beq{\begin{equation}}
\def\enq{\end{equation}}

\def\jcap{Jour. Cosmology and Astro-Particle Phys.\,}
\def\mnras{M.N.R.A.S.\,}

\def\apj{Astrophys.J.\,}
\def\apjl{Astrophys.J.Lett.\,}

\def\nat{Nature\,}

\def\prl{Phys. Rev. Lett.\,}

\def\prd{Phys.Rev.D\,}
\def\araa{Annu.Rev.Astron.Astrophys.\,}
\def\aap{Astron.Astrophys.\,}
\def\aaps{Astron.Astrophys.Supp.\,}

\def\physrep{Phys.Rep.\,}

\def\bec{\begin{center}}
\def\enc{\end{center}}

\def\fermi{\textit{Fermi~}}

\begin{document}

  \title[Gamma-Ray Bursts]
    {Relativistic Astrophysics}

  \author{P. M\'esz\'aros and M.J. Rees}




\pagenumbering{arabic}
\setcounter{page}{1}


\setcounter{figure}{0}

\centerline{\bf Gamma-Ray Bursts}
\centerline{P. M\'esz\'aros$^1$, M.J. Rees$^2$}
\noindent
$^1${Dpt. of Astronomy and Astrophysics; Dpt. of Physics;
Ctr. for Particle and Gravitational Astrophysics; Inst. for Gravitation and the Cosmos;
Pennsylvania State University, 525 Davey Lab, University Park, PA 16802, USA}\\
\noindent
$^2${Institute of Astronomy, Cambridge University, Madingley Road, Cambridge CB30HA, U.K.} 

\section{Historical Overview and Basic Concepts}
\label{sec:intro}

Gamma-Ray Bursts (GRBs) were serendipitously discovered in the late 1960s by the military 
Vela satellites which were monitoring the  Nuclear Test Ban Treaty between the US and the 
Soviet Union. The announcement was postponed for several years, after having ruled out
a man-made origin and ascertained that they were outside the immediate solar system 
\cite{Klebesadel+73}. In a matter of a few years more than a hundred models had been
proposed to explain their astrophysical origin \cite{Ruderman75grb}, ranging from comet
infalls, through stellar cataclysmic events, to events associated with supermassive black 
holes at the center of galaxies. The problem in making the first steps towards a theoretical 
understanding was that the gamma-ray instruments of the time had poor positional accuracy, 
transmitted to Earth only many hours after the trigger, so that only wide-field, insensitive 
telescopes could follow-up the bursts to look for counterparts at other wavelengths.   

In the 1990s the Compton Gamma Ray Observatory (CGRO) was launched, one of whose main
objectives was the detection of GRBs. The Burst and Transient Source Experiment (BATSE) onboard 
CGRO obtained, over a decade, the positions of $\sim$ 3000 GRBs. This showed that they were 
uniformly distributed over the sky \cite{Meegan+92isotropy}, indicating either an extragalactic 
or a `galactic-halo origin.  BATSE also found that GRBs can be classified into two duration classes, 
short and long GRBs, with a dividing line at $\sim 2$ s \cite{Kouveliotou+93}. 

The search for GRB counterparts at other wavelengths remained unsuccessful for almost 25 years,
until in 1997 the Beppo-SAX satellite localized with greater accuracy the first long lasting X-ray
afterglows \cite{Costa+97}, which in turn enabled the first optical host galaxy identification
and redshift measurement \cite{Vanparadijs+97}. 
The long bursts were found to be associated with galaxies where active star formation was
taking place, typically at redshifts $z \sim 1-2$, and in some cases a supernova of type
Ic was detected associated with the bursts, confirming the stellar origin of this class.
The power law time decay of the light curve was also observed, in a number of cases, to exhibit
a steepening after $\sim 0.5-1$ day, suggesting (for reasons explained below) 
that the emission was collimated into a jet, 
of typical opening half-angle $\sim 5^o$, which eased the energy requirements. Even so, 
at cosmological distances this implied a total time-integrated energy output of 
$\sim 10^{50}-10^{51}$ erg. This is roughly $10^{-3}$ of a solar rest mass, emitted over
tens of seconds. This is more than our Sun emits over its ten billion year lifetime, and
about as much as the entire Milky Way emits over a hundred years --
and that is mainly concentrated into gamma rays.

Well before the CGRO and Beppo-SAX observations, early theoretical ideas about the origin of 
GRBs had converged towards an energy source provided by the gravitational potential of a compact 
stellar source, the latter being suggested by the short duration (tens of seconds)  and
fast variability $\gtrsim 10^{-3}$ s of the $\gamma$-ray emission, using a simple causality argument 
$R \lesssim  c\Delta t \lesssim$ 10-100 Km. 
The large energies liberated in  a small volume and in a short time, as well as the
observed hard spectrum ($\gtrsim$ MeV) would then produce 
abundant electron-positron pairs via photon-photon interactions, creating a hot fireball 
which would expand, eventually reaching relativistic bulk velocities \cite{Cavallo+78}. 

Among the first  stellar sources discussed which could be responsible for GRBs were binary
double neutron star (DNS) mergers, or black hole-neutron star (BH-NS) mergers, whose occurrence 
rate as well as the expected energy liberated $\sim GM^2/R$ appeared sufficient for powering 
even extragalactic GRBs \cite{Eichler+89ns,Paczynski90wind,Meszaros+92tidal,Narayan+92merg}. 
These are nowadays, the leading candidates for the short gamma-ray bursts, as shown
by Swift and other observations e.g.  \cite{Gehrels+09araa}. 
Another candidate stellar source was the core collapse of massive stars and 
the accretion into the resulting black hole  \cite{Woosley93col,Paczynski98grbhn}.
Initially it was thought that this would result in a GRB and a failed supernova, but later
observations, e.g.  \cite{Galama+98-980425} and others, showed an unusually luminous  
core collapse supernova of type Ic associated  with some GRBs; these supernovae have
since been referred to as hypernovae. The core collapse model, referred to as a collapsar, 
is currently well established as the source of most long GRBs.

The predicted rate of occurrence of binary mergers and of hypernovae is sufficient to 
account for the number of bursts observed, even if the gamma-rays are beamed to the 
extent that only one event in 100-1000 is observed. (We expect less than one observable 
burst  per million  years from a typical galaxy, but the detection rate can nonetheless 
be of order one per day because that are so powerful that they can be detected out to 
the Hubble radius)

\section{CGRO Results and Basic Models}
\label{sec:cgro}

The dynamics of the expected relativistic fireball expansion were investigated by 
\cite{Paczynski86,Goodman86grb}. The fact that photons  of over  100 MeV are detected 
provides compelling evidence for ultra-relativistic expansion. To avoid degradation 
of the spectrum  via photon-photon interactions to energies  below the electron-positron 
formation threshold $m_e c^2=$ 0.511 MeV the outward flow must have a bulk Lorentz factor  
$\Gamma$ high enough so that the relative angle at which the photons collide is less 
than $\Gamma^{-1}$, thus diminishing the pair production threshold
\cite{Fenimore+93higammma,Harding+94higamma}.

Since each baryon in the outflow must be given an energy exceeding 100 times its rest mass, 
a key requirement of the central engine is that it must concentrate a lot of its energy into 
a very small fraction of its total mass. This favours models were magnetic fields and 
Poynting flux are important.

The observed spectrum extends to high energies,  generally in a broken power law shape, 
i.e., highly nonthermal.  Two initial problems \cite{Paczynski90wind,Shemi+90}
with the first expanding fireball models 
were that (a) they are initially optically thick and  the photon spectrum escaping from the 
Thompson scattering photosphere would be expected to be an approximate blackbody, 
and (b) most of the initial fireball energy would be converted into kinetic energy of expansion,
with a concomitantly reduced energy in the observed photons, i.e. a very low radiative efficiency.
\begin{figure}[htb]
\includegraphics[width=1.0\textwidth,height=2.5in,angle=0.0]{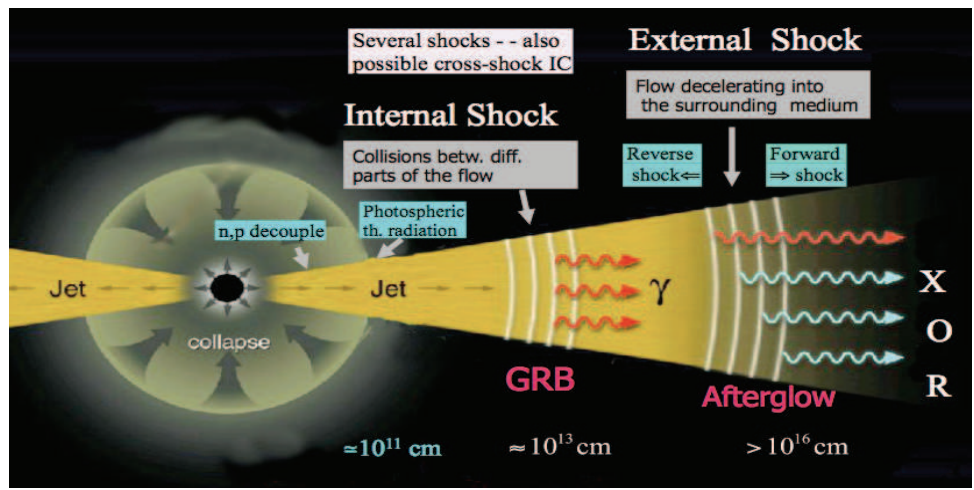}
\caption{Schematic GRB jet from a collapsing star.}
\label{fig:jet-schem}
\end{figure}

A simple way to achieve a high efficiency and a nonthermal spectrum, which is currently
the most widely invoked explanation, is by reconverting the kinetic energy of the flow 
into random energy via shocks, after the flow has become optically thin \cite{Rees+92fball}.
Two different types of shocks may be expected. There will be an  external shock, when the 
expanding fireball runs into the external interstellar medium or a pre-ejected stellar wind,
and a reverse shock propagating back into the ejecta. As in supernova remnants, 
Fermi acceleration of electrons into a relativistic power distribution in the turbulent magnetic
fields boosted in the shock leads to synchrotron emission \cite{Rees+92fball,Meszaros+93impact}
resulting in a broken power law spectrum, where the high energy photon spectral slope fits easily
the observations, and the single electron low energy photon slope -2/3 can, with a distribution 
of minimum energy electrons $\gamma_{min}$, reproduce the observed average low energy photon 
slope values of -1 (see also \cite{Katz94spectra,Sari+95hydro}). The reverse shock
would lead to optical photons, while inverse Compton emission in the forward blast wave would
produce photons in the GeV-TeV range \cite{Meszaros+93multi}.  
\begin{figure}[htb]
\includegraphics[width=1.0\textwidth,height=3.0in,angle=0.0]{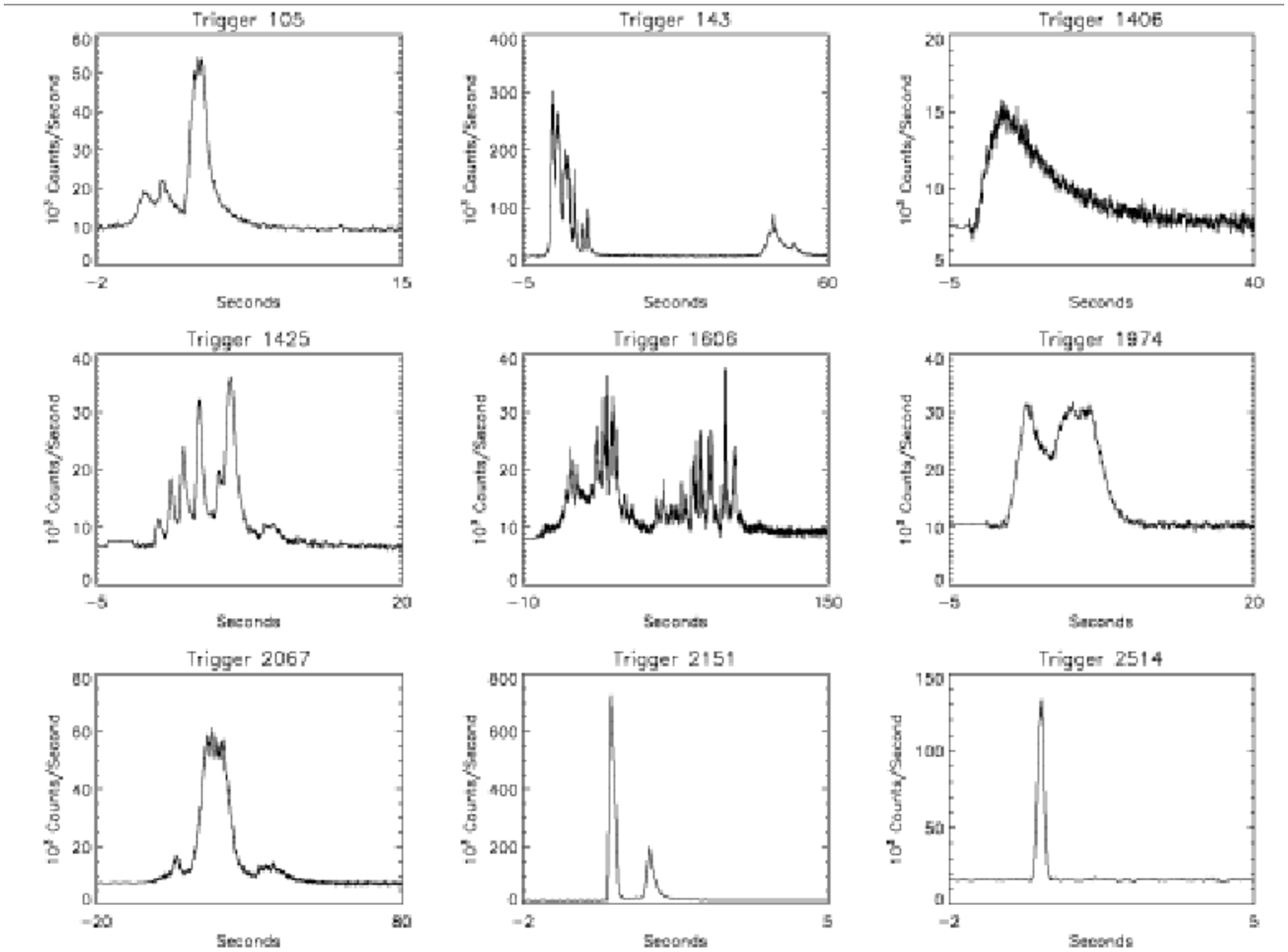}
\caption{A diversity of gamma-ray light curves from the BATSE instrument on the Compton 
Gamma Ray Observatory.}
\label{fig:batse-lc9}
\end{figure}

There could, additionally, be dissipation and acceleration within the outflowing jet itself. 
If the jet is unsteady,  internal shocks \cite{Rees+94unsteady,Paczynski+94is} can form 
as faster portions of the flow catch up with slower portions. 
And if magnetic stresses are important within the jets (i.e. they are  Poynting dominated 
outflows \cite{Meszaros+97poynt}, instead of the usual baryonic inertia dominated 
outflows) then magnetic reconnection can provide efficient mechanical conversion of bulk 
into random energy \cite{Lyutikov+03grbmag} (see also \cite{Usov94,Thompson94})  Any of these 
models provide a generic scenario  for explaining the radiation spectrum, largely  independent 
of the specific nature of the progenitor.
 
Internal shocks continue to be the model most widely used by observers to interpret the prompt 
MeV emission, while the external shock model is the favored interpretation for the long-term 
afterglows starting at high energies and phasing into gradually longer wavelengths over periods of
days to months. Coincidentally, the detection of the afterglows was preceded, a few weeks earlier,
by the publication of quantitative predictions of the power law spectral and time dependence 
of X-ray, optical and radio afterglows \cite{Meszaros+97ag}, in general agreement with observations.
Prompt optical afterglows were first detected in 1999 \cite{Akerlof+99},  while multi-GeV emission 
was reported by CGRO-EGRET \cite{Hurley+94egret}, and  more recently and in greater detail
by Fermi (see \S \ref{sec:fermi}).

\section{Beppo-SAX and HETE-2 Results and Issues}
\label{sec:var}

The evidence for a jet outflow is based on the observed steepening of the light curve
after $\sim$ day \cite{Frail+01beam}, 
which is attributed to the transition between the afterglow early relativistic 
expansion, when the light-cone is narrower than the jet opening half-angle $\theta_j$
and the late expansion, when the light-cone has become wider than the jet, $\Gamma^{-1}
\geq \theta_j$, leading to a drop in the effective flux \cite{Rhoads97jet,Kulkarni+99jet,
Meszaros+99rev}. A jet opening half-angle $\theta_j\sim$ 3-5 degrees is inferred, which 
reduces the total energy requirements to about $10^{51}-10^{52}$ ergs. This,  even
allowing for substantial inefficiencies, is compatible with currently favored scenarios 
based on a stellar collapse or a compact merger, e.g. \cite{Gehrels+09araa} and 
\S \ref{sec:intro}.

Observations with
the Beppo-SAX and HETE-2 satellites indicated the existence of a sub-class of GRBs called 
X-ray flashes (XRFs), whose spectrum peaks at energies 30-80 keV instead of the 300 keV - 1 MeV 
of classical GRBs, and with wider jet opening angles, e.g. \cite{Dalessio+06xrf}. 
The relative frequencies  of XRFs versus GRBs led  to considerations about a possible 
continuum distribution of angles, as well as about the jet angular shape, including 
departures from simple top-hat (abrupt cut-off) including an inverse power law or a Gaussian 
dependence on the angle \cite{Rossi+02struct,Kumar+03struct,Zhang+04unijet,Lamb+05jet}.

A problem with simple internal shock synchrotron models of the prompt MeV emission is that
the low energy photon number spectral slope, which is expected to be -2/3, is found to be
flatter in a fraction of BATSE bursts \cite{Preece+98death}. In addition, the synchrotron 
cooling time can be typically shorter than the dynamical time, which would lead to slopes 
-3/2 \cite{Ghisellini+99grbspec}. In either internal shock Fermi acceleration or in magnetic 
reconnection schemes, a number of effects can modify the simple synchrotron spectrum to
satisfy these constraints. Another solution involves a photospheric component, discussed below.

A natural question is whether the clustering of spectral peak energies in the 0.1-0.5 MeV range 
is intrinsic or due to observational selection effects \cite{Preece+98peak,Dermer+99extshock}. 
A preferred peak energy may be attributed to a blackbody spectrum at the comoving pair 
recombination temperature in the fireball photosphere \cite{Eichler+00thermal}. 
A photospheric component can address also the above low-energy spectral slope issue with
its steep Rayleigh-Jeans part of the spectrum, at the expense of the high energy power law. 
This was generalized \cite{Meszaros+00phot} to a photospheric blackbody spectrum at low energies 
with a comptonized photospheric component and possibly an internal shock or other dissipation 
region outside it producing Fermi accelerated electrons and synchrotron photons at high energies.
Photospheric models with moderate scattering depth can in fact lead to a Compton equilibrium 
which gives spectral peaks in the right energy range \cite{Peer+04waxb} and positive low energy
slopes as well as high energy power law slopes (the positive low energy slopes can always be 
flattened through a distribution of peak energies). A high radiative efficiency can be a problem
if the photosphere  occurs beyond the saturation radius $r_{sat}\sim r_0 \eta$, where $r_0$
is the base of the outflow and $\eta=L/{\dot M}c^2$ is the asymptotic bulk Lorentz factor
\cite{Meszaros+00phot}. However, a high radiation efficiency  with low and high energy slopes
can be obtained  in all cases if significant dissipation (either magnetic reconnection or shocks) 
is present in the photosphere \cite{Rees+05phot,Peer+06phot}. This can also address the 
phenomenological Amati \cite{Amati+02episo} and Ghirlanda \cite{Ghirlanda+04-EpkEjet} relations 
between spectral peak energy and burst fluence \cite{Rees+05phot,Thompson+07phot}

\section{Bursts in the Swift Era}
\label{sec:swift}

The launch of the Swift satellite in 2004 ushered in a new era of extensive data 
collection and analysis on GRBs, at wavelengths ranging from optical to MeV energies. 
This resulted in a number of interesting new discoveries, which have motivated various
refinements and reappraisals as well as new work on theoretical models, as discussed 
at greater length in the next sections.

Swift is equipped with three instruments: the Burst Alert Telescope (BAT), the X-Ray 
Telescope (XRT) and the UV Optical Telescope (UVOT).  The BAT detects bursts and locates them 
to about 2 arcminutes accuracy. This position is then used to automatically slew the spacecraft,
typically within less than a minute,  re-pointing the high angular resolution XRT and UVOT 
instruments towards the event.  The positions are also rapidly sent to Earth so that ground 
telescopes can follow the afterglows.  

A surprising new result achieved by Swift was that in a large fraction of the bursts
the X-ray afterglow shows an initial very steep time decay, starting after the end of the prompt
$\gamma$-ray emission. This then is generally followed by a much shallower time decay, often
punctuated by abrupt, large amplitude X-ray flares, lasting sometimes for up to $\sim$ 1000 s,
which then steepens into a power law time decay with the more usual (pre-Swift) slope
of index of roughly -1.2 to -1.7 \cite{Zhang+06ag,Nousek+06ag}. A final further steepening is
sometimes detected, ascribed to beaming due to a finite jet opening angle.
The initial steep decay may be ascribed to the evanescent radiation from high latitudes 
$\theta>\Gamma^{-1}$ relative to the line of sight \cite{Kumar+00naked,Zhang+09steepdec}, 
while the ensuing shallow decay phase may be due to continued outflow of material after the
prompt emission has ended \cite{Liang+07shallow}, which may undergo occasional internal shocks 
resulting in X-ray flares, e.g. \cite{Zhang+06ag,Falcone+07flare,Gehrels+09araa}. The subsequent 
steepening can be ascribed to the previously known forward shock gradual deceleration and the 
beaming induced jet break. These structures in the X-ray afterglow light curves are present both
in long and short bursts.
\begin{figure}[htb]
\includegraphics[width=1.0\textwidth,height=3.0in,angle=0.0]{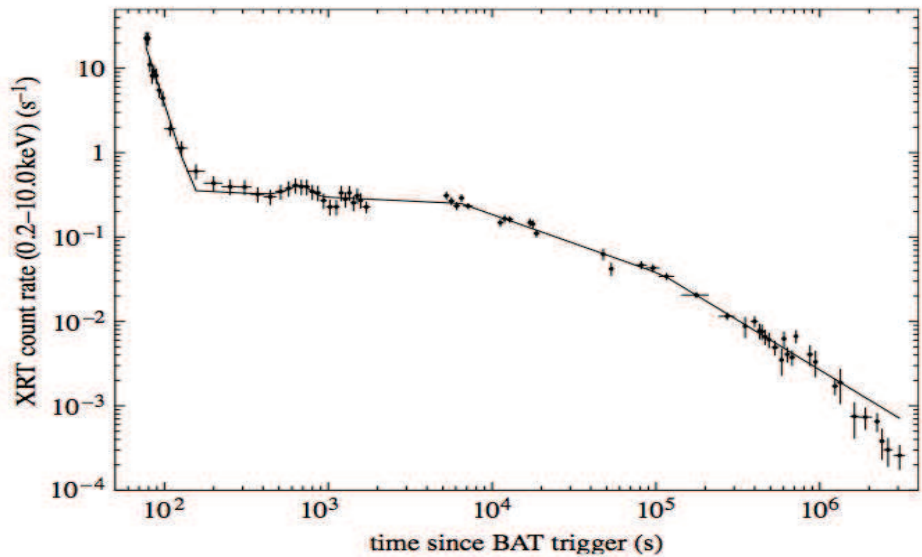}
\caption{Light curves of GRB060428A from the Swift XRT \cite{Burrows+07xrf}.}
\label{fig:GRB060428A-xrlc}
\end{figure}

Long GRBs (LGRBs) are found in galaxies where massive stars are forming, present over a large 
redshift range from $z=0.0085$ to $z>8$. Most LGRBs that occur near enough for supernova 
detection have an accompanying Type Ib or Ic supernovae, supporting the growing evidence that 
LGRBs are caused by ``collapsars" where the central core of a massive star collapses to a 
compact object such as a black hole or possibly a magnetar.

The number of GRB redshifts obtained underwent a rapid expansion after the launch of
Swift (currently in excess of 200), thanks to the rapid localization allowing large 
grounds-based telescopes to acquire high quality spectra while the afterglow was still bright.
The most distant ones are intrinsically the brightest, typically $E_{iso} \gtrsim 10^{55}$
erg, the current record holder being GRB090423 at a spectroscopically confirmed redshift 
$z=8.2$ \cite{Tanvir+09-z8.2}, and GRB090429B, at a photometric redshift $z\sim 9.4$
\cite{Cucchiara+11-z9}.

With the increasing statistics, LGRBs are contributing to a better understanding of the 
high- redshift universe. They provide spectroscopic information about the chemical composition 
of the intervening intergalactic medium at epochs when the Universe was as low as 1/20th of 
its present age.  Also, since LGRBs are the endpoints of the lives of massive stars, their 
rate is approximately proportional to the star formation rate. This gives information at 
high redshift where the rate is highly uncertain. There can be evolutionary biases, such as 
a dependence of LGRBs on the metalicity of host galaxies, which must be taken into account 
\cite{Kistler+09sfrgrb,Robertson+12sfrgrb}.

Swift succeeded in finally localizing the host galaxies of a number of short GRBs (SGRBs).
e.g. \cite{Berger+05-050724,Fox+05-050709}. Unlike long GRBs, the SGRBs typically originate 
in host galaxies with a wide range of star formation properties, including low formation rates. 
The host properties are substantially different than those of LGRBs \cite{Fong+13sgrbhost,
Fong+10sgrbhost,Leibler+10sgrbhost}, indicating a different origin. Furthermore, nearby SGRBs 
show no evidence for simultaneous supernovae \cite{Nakar07sgrb}, as do many long bursts. 
These results reinforce the interpretation that SGRBs arise from an old population of stars, 
probably due to mergers of compact binaries such as double neutron star or neutron star-black 
holes \cite{Nakar07sgrb,Eichler+89ns,Paczynski90wind}.

Short GRBs are found to have generally a lower isotropic-equivalent luminosity and total 
energy output $E_{iso}$ than LGRBs, typically $E_{iso}\sim 10^{50}$ ergs, with a weak afterglow,
and in the few cases where a jet break has been measured, the jet opening angle appears to be 
wider than in LGRBs, $\theta_j \sim 5^o -25^o$ \cite{Burrows+06break,Fong+12breaksho}.
Another new result was the discovery, in about 25\% of SGRBs, of a longer ($\sim 100$ s) 
light curve tail with a spectrum softer than the initial episode \cite{Norris+06sgrbtail,
Gehrels+06-060614}.  This is puzzling in the context of double neutron star or neutron 
star-black hole mergers, since numerical simulations suggest that the disk of disrupted 
matter is accreted in at most a few seconds, e.g. \cite{Nakar07sgrb}.  A longer accretion 
timescale, however, may occur if the disk is highly magnetized \cite{Proga+06flare}, or if 
the compact merger results in a temporary magnetar whose magnetic field holds back the 
accretions disk until the central object collapses to a black hole \cite{Rezzolla+11sgrb-mag}.

\subsection{Other types of bursts}
\label{sec:other}

The demarcation into two classes of bursts is however too simplistic to be the whole story. 
Some bursts fit neither category. For example,  some bursts detected by Swift are extreme 
magnetar flares – caused by the sudden readjustment (and release of stored energy) in 
the magnetosphere of a highly magnetized ($\gtrsim 10^{14}$ G) neutron star.   These are 
of interest for phenomenologists but are a confusing complication for those seeking 
correlations between the observable parameters of bursts.

But the Swift spacecraft has revealed another type of object that is of great interest, 
and which was a surprise:  bursts characterized by unusually persistent and prolonged 
emission, and located at the centre of the host galaxy. These are interesting both to 
astrophysicists and to relativists, as they may be triggered by a long-predicted effect 
that has not before been conclusively detected: the tidal disruption of a star by a 
massive hole. 

Tidal capture and disruption of stars attracted interest back in the 1970s, when theorists 
started to address the dynamics of stars  concentrated in a high-density `cusp' surrounding 
the kind of black hole expected to exist in the centres of galaxies (and perhaps in some 
globular star clusters as well). It was recognized   that stars could be captured and 
swallowed by the  central hole if they were in a `loss cone' of near-radial orbits. 

If the central hole is sufficiently massive, tidal forces at the horizon may be too gentle 
to disrupt to  star while it is still in view, in which case it is captured without any 
conspicuous display.  For a solar-type star, this requires $\sim 10^8 M_\odot$; for white 
dwarfs the corresponding mass is $\sim 10^4 M_\odot$. (And neutron stars are swallowed 
whole by black holes with masses above about $10 M_\odot$ -– this is important for the 
gravitational wave signal in coalescing binary stars, as discussed elsewhere in this volume). 
For a spinning 
hole, the cross-section for capture, and the tidal radius for disruption, depend on the 
relative orientation of the orbital and spin angular momenta. (Stars on orbits 
counter-rotating with respect to the hole are preferentially captured: this is a process 
that would reduce the spin of a hole in a galactic nucleus.)

When stars are swallowed before disruption, they can be treated as point mass particles 
moving in the gravitational field of the hole; their interactions among themselves can 
be treated the same way, except insofar as star-star collisions are important. But the 
physics is much messier in the cases when the tidal radius is outside the hole and the 
star is disrupted rather than swallowed whole. This phenomenon has been studied since 
the 1970s, first via analytic models (e.g. \cite{Carter+83tid,Rees88tid}) and  subsequently 
by progressively more powerful numerical simulations (e.g. \cite{Kobayashi+04disrupt,
Decolle+12tid}, etc.). In the Newtonian approximation the tidal radius is
$R_t \sim R_\ast \left({M_{BH}}/{M_\ast}\right)^{1/3}$.
There are several key parameters: the type of star; the pericentre of the star's orbit 
relative to the tidal radius, and the orientation of the orbit relative to the hole's 
spin axis. In most astrophysical contexts, the captured stars would be on highly 
eccentric orbits (i.e the orbital binding energy would be small compared to that of a 
circular orbit at the tidal radius). If the pericentre is of order $R_t$, the star will 
be disrupted, and the debris will be continue on eccentric orbits, but with a spread of 
energies of order the binding energy of the original star. Indeed nearly half  the 
debris will escape from the hole’s gravitational field completely; the rest will be on 
more tightly bound (but still eccentric ) orbits, and would be fated to dissipate further, 
forming a disc much of which would then be accreted into the hole.  A pericentre passage 
at (say) 2 or 3 times $R_t$ would not disrupt a star completely, but would remove its 
envelope, and induce internal oscillations, thereby extracting orbital energy and leaving 
the star vulnerable on further passages. On the other hand,  as first discussed by
\cite{Carter+83tid}, a star that penetrates far inside the tidal radius (but not 
so close to the hole that it spirals in) will be drastically distorted and compressed by 
the tidal forces, perhaps to the extent that a nuclear explosion occurs, leading to a 
greater spread in the energy of the debris than would result from straight gas dynamics.

There have in recent years been detailed computations of these processes, and also of the 
complicated and dissipative gas dynamics that leads to the accretion of the debris, and 
the decline of the associated luminosity as the dregs eventually drain away. There are 
two generic predictions: the debris enveloping the hole should initially have a thermal 
emission with a power comparable to the Eddington luminosity of the hole; and at late 
times, when the emission comes from the infall of debris  from orbits with large apocentre, 
the luminosity falls as $L\propto t^{-5/3}$.

 There has been much debate about the role of tidal capture in the growth of supermassive 
holes, and the fueling of AGN emission,  and many calculations of the expected rate, 
taking account of what has been learnt about the masses of holes, and the properties of the 
stellar populations surrounding them.  Some flares in otherwise quiescent galactic nuclei, 
where the X-ray luminosity  surges by a factor $\gtrsim 100$,  have been attributed to 
tidal disruptions.

 But tidal disruption is included in this chapter mainly  because of a remarkable burst 
detected by Swift, Sw J164449.3 \cite{Burrows+11-1644tid}, 
located at the centre of its host galaxy, and which was 
exceptionally prolonged in its emission. This is perhaps the best candidate so far for an 
event triggered by tidal capture of a star. The high energy radiation, were this model 
correct, would come from a jet generated near the hole.  Modeling is still tentative, 
and is difficult because there is no reason to expect alignment between the angular 
momentum vectors of the hole and of the infalling  material.  But the inner disc (and 
therefore the inner jet) would be expected to  align with the hole, though it is possible 
that the jet is deflected further out by material with different alignment 
(c.f. \cite{Mckinney+13mhdjet}).
  
Be that as it may, this exceptional burst offers model-builders an instructive 
`missing link' between the typical long (`Type 1') burst, involving a massive star, and 
the jets in AGNs which are generated by processes around supermassive holes.

\section{Bursts at Energies above GeV: Fermi and beyond}
\label{sec:fermi}

The Fermi satellite, launched in 2008, has two instruments: the Gamma-ray Burst Monitor (GBM, 
\cite{Meegan+09GBM}) and the Large Area Telescope (LAT, \cite{Atwood+09LAT}). The GBM measures 
the spectra of GRB in the energy range from 8 keV to 40 MeV, determining their position to 
$\sim 5^o$ accuracy. The LAT measures the spectra in the energy range from 20 MeV to 300 GeV,
locating the source positions to an accuracy of $< 1^o$.  The GBM detects GRBs at a rate of 
$\sim$ 250 per year, of which on average 20\% are short bursts, while the LAT detects bursts 
at a rate of $\sim 8$ per year.  The great strength of this combination is to provide the
large field of view and high detection rate of the GBM extending to energies as low as the 
BAT in Swift, with the very high energy window of the LAT, which opens up a whole new vista 
into the previously almost unexplored GeV to sub-TeV range of GRBs. 
%
%

Two unexpected features of the GeV emission of bursts were soon discovered by the Fermi-LAT. 
One is that the onset of the GeV emission is invariable delayed relative to the onset of
the MeV emission (by a few seconds in LGRBs, and fractions of a second in SGRBs), e.g.
\cite{Abdo+09-080916,Abdo+09-090510,Ackermann+10-090510,Abdo+09-090902}.
The other is that the GeV emission generally lasts for much longer then the MeV emission,
decaying as a power law in time and lasting up to a 1000 s in some cases, i.e. well into 
the afterglow phase, including both LGRBs and SGRBs. 
The fact that GeV emission has been detected from a number of
SGRBs is, in itself, also new. Remarkably, the GeV behavior of LGRBs and SGRBs is quite 
similar. This is not unexpected, since most of the GeV emission is produced in the afterglow
phase, which is essentially a self-similar process. What is more unexpected is that 
the ratio of the total energy in the GeV range to MeV range is $\sim 0.1 -0.5$ for LGRBs, 
while it is $\gtrsim 1$ for SGRBs.

%
\begin{figure}[htb]
\includegraphics[width=1.0\textwidth,height=2.5in,angle=0.0]{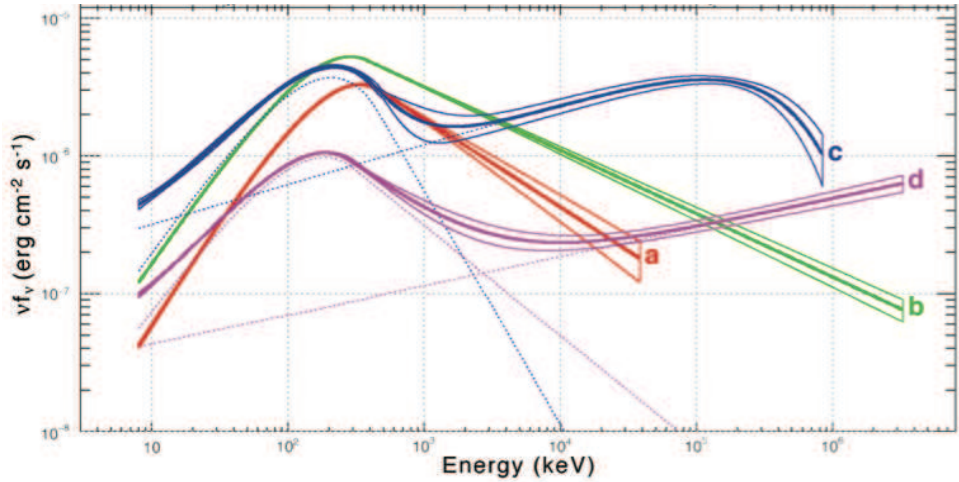}
\caption{Spectra of GRB090926A from \fermi at four different time intervals,
a= [0.0-3.3s], b= [3.3-9.7s], c= [9.7-10.5s], d= [10.5-21.6s]
\cite{Ackermann+11-090926}.}
\label{fig:GRB090926A-spec}
\end{figure}

Bursts detected with the LAT have spanned a range of redshifts extending up to $z=4.3$,
with photon energies (in the burst rest frame) up to $10-130$ GeV, the highest value so 
far being being that found for GRB 130427A \cite{130427aLAT}, at a redshift $z\sim 0.33$.
This is encouraging for the planned large Cherenkov Telescope Array (CTA) \cite{Bouvier+11-CTA,
Funk+12-CTA}, whose energy threshold may be as low as 50 GeV and whose detection rate of GRBs is 
estimated in the range $0.7-1.6$ per year, based on the rate of Swift triggers. A roughly
similar rate of detection is also expected for the High Altitude Water Cherenkov (HAWC)
detector \cite{Abeysekara+12hawc,Taboada+13hawc}, whose threshold is expected to be 10-20 GeV.

\section{GRBs in Non-photonic Channels?}
\label{sec:nonphot}

Two types of non-photonic signals that may be expected from GRBs are gravitational
waves (GWs) and high energy neutrinos (HENUs). The most likely GW emitters are short
GRBs \cite{Centrella+11gw-em}, if these indeed arise from  merging compact objects
\cite{Gehrels+09araa}. 
The Swift and Fermi localization of a short GRB would help to narrow the search window for 
gravitational waves from that object \cite{Finn+99-grbgw}. The detection of gravitational waves 
from a well-localized GRB would lead to a great scientific payoff for understanding the merger 
physics, the progenitor types, and the neutrons star equations of state.
The rates of compact merger GW events in the advanced LIGO and VIRGO detectors may be at least 
several per year \cite{Leonor+09-ligogrb}.  However, even if these events all give rise to 
gamma ray bursts, only a small fraction would be beamed towards us. 
Long GRBs, more speculatively, might be detectable
in GWs if they go through a magnetar phase \cite{Corsi+09mag}, or if the core collapse
breaks up into substantial blobs \cite{Kobayashi+03gwgrb}; more detailed numerical
calculations of collapsar (long) GRBs lead to GW prospects which range from
pessimistic \cite{Ott+11gwcoll} to modest \cite{Kiuchi+11gwcoll}.

High energy neutrinos may also be expected from baryon-loaded GRBs, if sufficient protons
are co-accelerated in the shocks. The most widely considered paradigm involves
proton acceleration and $p\gamma$ interactions in internal shocks, resulting in
prompt $\sim 100$ TeV HENUs \cite{Waxman+97grbnu,Murase+06grbnu}. Other interaction
regions considered are external shocks, with $p\gamma$ interactions on reverse shock UV
photons leading to EeV HENUs \cite{Waxman+00nuag}; and pre-emerging or choked jets
in collapsars resulting in HENU precursors \cite{Meszaros+01choked}.
An EeV neutrino flux is also expected from external shocks in very massive Pop. III
magnetically dominated GRBs \cite{Gao+11pop3nu}. Current IceCube observations
\cite{Abbasi+12grbnu-nat} are putting significant constraints on the simplest internal 
shock neutrino emission model. More careful modeling of internal shocks \cite{Hummer+11nu-ic3}
reveal that several years of observations will be needed for reliably testing such models,
while other types of models, such as photospheric models \cite{Gao+12photnu} or modified 
internal shock models \cite{Murase+12reac} are yet to be tested. However, the excitement in 
this field is  palpable, especially since the announcement of the detection by IceCube of 
PeV neutrinos \cite{IC3+13pevnu} whose origin is almost certainly astrophysical.

\bigskip
\noindent
{\it Acknowledgments:}
We are grateful to NASA NNX 13AH50G and the Royal Society (U.K.) for partial support.





%




\end{document}